\documentstyle[prl,aps]{revtex}

\begin{document}

\draft

\title{Comment on the paper ``Bound States in the One-dimensional
Hubbard Model''}

\author{F. H. L. E{\ss}ler}

\address{Department of Physics, Theoretical Physics, Oxford University,
1 Keble Road, Oxford OX1 NP, UK}

\author{F. G\"ohmann, V. E. Korepin}

\address{Institute for Theoretical Physics, State University of New
York at Stony Brook, Stony Brook, NY 11794-3840, USA}

\maketitle

\begin{abstract}
We comment on the preprint cond-mat/9805103 by D. Braak and N. Andrei
\cite{BrAn98}. We point out that the ``new'' Bethe Ansatz equations
presented in \cite{BrAn98} are identical to the Bethe equations for
strings introduced by M. Takahashi for the description of thermodynamics
in 1972 \cite{Takahashi72}. Some physics suggested in \cite{BrAn98}
is incorrect. In particular, all former conclusions made on the basis
of the string Bethe equations remain valid.  
\end{abstract}

(I) First let us correct a typo in \cite{BrAn98}. The bare $S$-matrix
for the scattering of an unbound particle with an $m$-complex (formula
(17) in \cite{BrAn98}) should be
\begin{equation}
     S_{k \phi_0^{(m)}}^{u (m)} = \frac{\sin k - \phi_0^{(m)} -
        m \text{i} u/4}{\sin k - \phi_0^{(m)} + m \text{i} u/4}
\end{equation}
(see eq.\ (6.59) of \cite{MuGo98a}). Let us also mention that by
convention the $S$-matrix in \cite{BrAn98} gets inverted upon
horizontal transposition of the indices. A straightforward calculation
shows that the ``new'' Bethe equations of \cite{BrAn98} ((BAC), eqs.\
(20)-(22) in \cite{BrAn98}) coincide with the standard string Bethe
equations \cite{Takahashi72}. This calculation involves adjustment of
notations and making explicit the so-called $\Lambda$-strings, which
are hidden in the notation of \cite{BrAn98} (see appendix A). [The
$\Lambda$ strings are configurations of the spin-rapidities
$\lambda_\gamma$ involving complex $\lambda_\gamma$. The
$\Lambda$-strings have to be distinguished from the so-called
$k$-$\Lambda$-strings, which are the subject of criticism in
\cite{BrAn98} ($k$-$\Lambda$-strings involve spin-rapidities
$\lambda_\gamma$ as well as charge rapidities $k_j$).] Since the
``new'' Bethe equations in \cite{BrAn98} coincide with the string
Bethe equations, they can not lead to new physics. It is no surprise,
in particular, that the counting of states mentioned in \cite{BrAn98}
gives the correct number $4^L$, since this counting is actually
identical to the calculation of \cite{EKS92b}.
The Algebraic Bethe Ansatz expression at the start of page 2 of
\cite{BrAn98} is incorrect: the Hamiltonian  does not act in this
way but needs to be replaced by the inhomogeneous transfer matrix of
the spin problem.

(II) Let us now demonstrate that the ``new physics'' discussed at the
end of [1] is partially incorrect and partially known. Let us
start with an incorrect statement of \cite{BrAn98}: in the last part
the holon-antiholon excitation for the half-filled repulsive Hubbard
model is considered. This excitation is known in the literature as
``charge singlet'' (we will follow the discussion given in
\cite{EsKo94a,EsKo94b}). It is described by two holes in the $k$-Fermi
sea ($k_1^h$ and $k_2^h$) and one $k$-$\Lambda$ string. The spin
parameter of the $k$-$\Lambda$ string is denoted by $\phi(q)$ in
\cite{BrAn98} and by $\Lambda'$ in \cite{EsKo94a,EsKo94b}. The authors
of \cite{BrAn98} state that the relation 
\begin{equation} \label{centre}
     2 \Lambda' = \sin k_1^h + \sin k_2^h
\end{equation}
is not valid (remember that $\Lambda' = \phi(q)$). This  statement is
incorrect: It was shown in \cite{EsKo94a,EsKo94b} that (\ref{centre})
follows from the string Bethe equations (which are identical to
(20)-(22) alias (BAC) in \cite{BrAn98}). Independent confirmation can
be found in \cite{klumper}. We repeat the derivation of
(\ref{centre}) in detail in appendix A below: Eq.\ (\ref{centre})
holds at half-filling! This confirms the expression for the $S$-matrix
obtained in \cite{EsKo94a,EsKo94b}. 

Let us comment on the ``new'' excitation below half-filling, which
is discussed in the last part of \cite{BrAn98}. Our comments are: (i)
the existence of this excitation was well known \cite{Takahashi72} and
(ii) is of no particular physical significance, since it (a) has a gap
and (b) is only one of an infinite number of gapful excitations below
half-filling. It is correct that eq.\ (\ref{centre}) is not satisfied
below half-filling and that as a result the $k$-$\Lambda$ string is an
independent excitation. However the same holds for the infinite number
of longer $k$-$\Lambda$ strings and the one-particle excitations!
These excitations also exist at zero density (zero filling)
\cite{MuGo98a}.  It is the generic situation for Bethe Ansatz solvable
models that the structure of the excitations persists, when going from
a trivial to a non-trivial ground state by changing the density (more
generally, an order parameter). This means in case of the Hubbard
model, that all strings as well as the one-particle excitations are
present at finite density below half-filling. Only their dispersion
curves get modified.  Similarly, for the XXX spin chain the string
excitations over the ferromagnetic ground state survive, when the
 magnetization  is diminished. Only the chain with zero magnetization
has a different excitation structure. Let us  emphasize  that the
$k$-$\Lambda$ string below half filling has a gap! 
Hence, the $k$-$\Lambda$ string does not influence the conformal
dimensions \cite{FrKo90} and does not lead to new low energy
physics. The same statement holds true for all longer $k$-$\Lambda$
strings. 

(III) Next, we want to comment on eq.\ (7) of \cite{BrAn98}.
The authors of \cite{BrAn98} do not require (7). They claim that this
is the reason for ``new'' physics. All new physics they present is
related to states containing one $k$-$\Lambda$ string (and several
real $k_j$ and real $\lambda_\gamma$). The authors of \cite{BrAn98} do
not argue the validity of their eqs.\ (5), (6) and (8). We show in
appendix B that for one $k$-$\Lambda$ string (and several real $k_j$
and real $\lambda_\gamma$) eq.\ (7) follows from (5), (6) and (8). This
shows an inconsistency in the arguments of \cite{BrAn98} and removes the
basis for the modification of the physics of the one-dimensional
Hubbard model.

(IV) Finally, let us comment on the continuity argument on the bottom
of the left column of the page 3 of \cite{BrAn98}. The defect of the
argument is the following: in the limit $u\to\infty$ [when the bound
state (9) satisfy periodic boundary condition] the energy of the bound
state is infinite, and it drops out of the Hilbert space. Continuity
to finite $u$ cannot be employed to count such states.

We would like to thank H. Frahm for helpful communications.

\appendix
\section{Derivation of equation (2)}
In this appendix we show that eqs.\ (20)-(22) in \cite{BrAn98} are
identical with the string Bethe equations of Takahashi and that
(\ref{centre}) follows from these equations.

Let us first cite the main results of \cite{BrAn98}: The $n$-complex
is  parameterized by an $n$-string of the form
\begin{equation}
     \phi_{a,j}^{(n)} = \phi_a^{(n)} + (n + 1 - 2j) \frac{\text{i} u}{4}
        \quad, \quad j = 1, \dots, n .
\end{equation}
[Note the slight change of notation compared to \cite{BrAn98}: We made
the replacement $\phi_0^{(n)} \rightarrow \phi_a^{(n)}$ (and
$\phi_j^{(n)} \rightarrow \phi_{a,j}^{(n)}$) which allows us to include
scattering of two different $n$ complexes of the same length. $a$
enumerates the $n$-strings.] The S-matrix of an unbound particle
with an $n$-complex is
\begin{equation}
     S_{k \phi_a^{(n)}}^{u (n)} = \frac{\sin k - \phi_a^{(n)} - 
        n \textstyle{\frac{\text{i} u}{4}}}{\sin k - \phi_a^{(n)} +
        n \textstyle{\frac{\text{i} u}{4}}},
\end{equation}
and the S-matrix of an $m$-complex with an $n$-complex is
\begin{equation}
     S_{\phi_b^{(m)} \phi_a^{(n)}}^{(m) (n)}
        \: = \: \frac{\phi_b^{(m)} - \phi_a^{(n)} -
	   |n - m| \textstyle{\frac{\text{i} u}{4}}}
          {\phi_b^{(m)} - \phi_a^{(n)} +
	   |n - m| \textstyle{\frac{\text{i} u}{4}}} \:
          \frac{\phi_b^{(m)} - \phi_a^{(n)} -
	   (n + m) \textstyle{\frac{\text{i} u}{4}}}
          {\phi_b^{(m)} - \phi_a^{(n)} +
	   (n + m) \textstyle{\frac{\text{i} u}{4}}} \:
	  \prod_{l=1}^{\min(m,n) - 1} \left(
          \frac{\phi_b^{(m)} - \phi_a^{(n)} -
	   (|n - m| + 2l) \textstyle{\frac{\text{i} u}{4}}}
          {\phi_b^{(m)} - \phi_a^{(n)} +
	   (|n - m| + 2l) \textstyle{\frac{\text{i} u}{4}}} \right)^2.
\end{equation}
Diagonalizing the transfer matrix leads to the following set of
equations
\begin{eqnarray} \label{bac1}
     e^{\text{i} k_j L} & = & \prod_{\delta = 1}^{M^u}
          \frac{\lambda_\delta - \sin k_j - \textstyle{
	  \frac{\text{i} u}{4}}}{\lambda_\delta - \sin k_j +
	  \textstyle{\frac{\text{i} u}{4}}}
          \prod_{(n,a)} S_{\phi_a^{(n)} k_j}^{(n) u}, \\ \label{bac2}
     \prod_{\delta \ne \lambda}^{M^u}
          \frac{\lambda_\gamma - \lambda_\delta -
	  \textstyle{\frac{\text{i} u}{2}}}{\lambda_\gamma -
	  \lambda_\delta + \textstyle{\frac{\text{i} u}{2}}}
	       & = &
     \prod_{j = 1}^{N^u}
          \frac{\lambda_\gamma - \sin k_j -
	  \textstyle{\frac{\text{i} u}{4}}}{\lambda_\gamma -
	  \sin k_j + \textstyle{\frac{\text{i} u}{4}}},
	       \\ \label{bac3}
     e^{\text{i} q^{(n)}(\phi_a^{(n)})L} & = &
          \prod_{(m,b) \ne (n,a)} S_{\phi_b^{(m)}
	  \phi_a^{(n)}}^{(m) (n)}
          \prod_{j=1}^{N^u} S_{k_j \phi_a^{(n)}}^{u (n)},
\end{eqnarray}
where
\begin{equation}
     q^{(n)} (\phi) = - 2 \text{Re} \arcsin(\phi + n \text{i} u/4) 
                      + (n + 1) \pi =
        - (\arcsin(\phi + n \text{i} u/4)
           + \arcsin(\phi - n \text{i} u/4)) + (n + 1) \pi
\end{equation}
and
\begin{equation}
     S_{\phi_a^{(n)} k_j}^{(n) u} =
          \left( S_{k_j \phi_a^{(n)}}^{u (n)} \right)^{-1}.
\end{equation}

Eqs. (\ref{bac1})-(\ref{bac3}) are the ``new'' equations of Braak and
Andrei ((20)-(22) or (BAC) in \cite{BrAn98}). In order to show that
(\ref{bac1})-(\ref{bac3}) agree with Takahashi's string Bethe equations
let us now adjust notations. Let
\begin{equation}
     U = u/4 \quad, \quad {\Lambda'}_a^m = \phi_a^{(m)} \quad,
     \quad \alpha = a \quad, \quad \beta = b.
\end{equation}
Let us further introduce the functions \cite{Takahashi72}
\begin{eqnarray}
     e(x) & = & \frac{x + \text{i}}{x - \text{i}} \\
     E_{nm} (x) & = & \left\{ \begin{array}{l} 
                      {\displaystyle
                      e \left( \frac{x}{|n - m|} \right)
                      e^2 \left( \frac{x}{|n - m| + 2} \right)
		      \cdots
                      e^2 \left( \frac{x}{n + m - 2} \right)
                      e \left( \frac{x}{n + m} \right)
		      \: \text{for} \quad n \ne m,} \\[3ex]
                      {\displaystyle
                      e^2 \left( \frac{x}{2} \right)
                      e^2 \left( \frac{x}{4} \right)
		      \cdots
                      e^2 \left( \frac{x}{2n - 2} \right)
                      e \left( \frac{x}{2n} \right)
		      \: \text{for} \quad n = m.}
		      \end{array} \right.
\end{eqnarray}
Then eqs.\ (\ref{bac1})-(\ref{bac3}) turn into
\begin{eqnarray} \label{tbac1}
     e^{\text{i} k_j L} & = & \prod_{\delta = 1}^{M^u}
	  e \left( \frac{\sin k_j - \lambda_\delta}{U} \right)
          \prod_{(n,\alpha)}
	  e \left( \frac{\sin k_j - {\Lambda'}_\alpha^n}{nU} \right),
	  \\ \label{tbac2}
     \prod_{j = 1}^{N^u}
          e \left( \frac{\lambda_\gamma - \sin k_j}{U} \right)
	       & = &
     - \prod_{\delta = 1}^{M^u}
          e \left( \frac{\lambda_\gamma - \lambda_\delta}{2U} \right),
	       \\ \label{tbac3} \nonumber
     \exp \{ - \text{i} L
          [\arcsin({\Lambda'}_\alpha^n + n \text{i} U) & + &
	  \arcsin({\Lambda'}_\alpha^n - n \text{i} U) +
	  (n + 1) \pi] \} \\ & = &
          - \prod_{j=1}^{N^u}
	  e \left( \frac{{\Lambda'}_\alpha^n - \sin k_j}{nU} \right)
          \prod_{(m,\beta)}
	  E_{nm} \left(
	  \frac{{\Lambda'}_\alpha^n - {\Lambda'}_\beta^m}{U} \right).
\end{eqnarray}
The spin rapidities $\lambda_\gamma$ in (\ref{tbac1}) and (\ref{tbac2})
may generally be complex. In order to obtain a set of equations which
contains only real unknowns and which transforms into a set of linear
integral equations in the thermodynamic limit we have to employ
Takahashi's string hypothesis for $\Lambda$ strings: As the number $N$
of electrons becomes large the spin rapidities are driven to string
positions characterized by their length $n$ and their real center
$\Lambda_\alpha^n$. Following Takahashi \cite{Takahashi72} we will
use the notation $\Lambda_\alpha^{n,j}$ instead of $\lambda_\gamma$.
$\Lambda_\alpha^{n,j}$ is the $j$-th spin rapidity involved in an
$n$-$\Lambda$ string with center $\Lambda_\alpha^n$,
\begin{equation}
     \Lambda_\alpha^{n,j} = \Lambda_\alpha^n + (n + 1 - 2j) \text{i}U.
\end{equation}
Following again Takahashi let us assume that in the thermodynamic
limit all $\lambda_\gamma$ are grouped into strings with an accuracy
of ${\cal O}(\exp( - \delta N))$, where $\delta$ is some positive
number. Then eqs.\ (\ref{tbac1})-(\ref{tbac3}) lead to
\begin{eqnarray} \label{ttbac1}
     e^{\text{i} k_j L} & = & \prod_{(n,\alpha)}
	  e \left( \frac{\sin k_j - \Lambda_\alpha^n}{nU} \right)
          \prod_{(n,\alpha)}
	  e \left( \frac{\sin k_j - {\Lambda'}_\alpha^n}{nU} \right),
	  \\ \label{ttac2}
     \prod_{j = 1}^{N^u}
          e \left( \frac{\Lambda_\alpha^n - \sin k_j}{nU} \right)
	       & = &
     - \prod_{(m,\beta)}
          E_{nm} \left( \frac{\Lambda_\alpha^n - \Lambda_\beta^m}{U}
	  \right), \\ \label{ttbac3} \nonumber
     \exp \{ - \text{i} L
          [\arcsin({\Lambda'}_\alpha^n + n \text{i} U) & + &
	  \arcsin({\Lambda'}_\alpha^n - n \text{i} U) +
	  (n + 1) \pi] \} \\ & = &
          - \prod_{j=1}^{N^u}
	  e \left( \frac{{\Lambda'}_\alpha^n - \sin k_j}{nU} \right)
          \prod_{(m,\beta)}
	  E_{nm} \left(
	  \frac{{\Lambda'}_\alpha^n - {\Lambda'}_\beta^m}{U} \right).
\end{eqnarray}

Taking logarithms we arrive at the following form of the string
Bethe equations, which is suitable for considering the thermodynamic
limit,
\begin{eqnarray} \label{t1}
     k_j L & = & 2 \pi I_j - \sum_{n=1}^\infty \sum_{\alpha = 1}^{M_n}
                 \theta \left(
		 \frac{\sin k_j - \Lambda_\alpha^n}{nU} \right)
                 - \sum_{n=1}^\infty \sum_{\alpha = 1}^{M_n'}
                 \theta \left(
		 \frac{\sin k_j - {\Lambda'}_\alpha^n}{nU} \right),
		 \\ \label{t2}
     \sum_{j=1}^{N - 2M'} \theta \left(
		 \frac{\Lambda_\alpha^n - \sin k_j}{nU} \right) & = &
		 2 \pi J_\alpha^n +
		 \sum_{m=1}^\infty \sum_{\beta = 1}^{M_m}
		 \Theta_{nm} \left(
		 \frac{\Lambda_\alpha^n - \Lambda_\beta^m}{U} \right),
		 \\ \label{t3}
     L [\arcsin({\Lambda'}_\alpha^n + \text{i}nU)
        + \arcsin({\Lambda'}_\alpha^n - \text{i}nU)] & = &
	         2 \pi {J'}_\alpha^n +
		 \sum_{j=1}^{N - 2M'} \theta \left(
		 \frac{{\Lambda'}_\alpha^n - \sin k_j}{nU} \right) +
		 \sum_{m=1}^\infty \sum_{\beta = 1}^{M_m'}
		 \Theta_{nm} \left(
		 \frac{{\Lambda'}_\alpha^n - {\Lambda'}_\beta^m}{U}
		 \right).
\end{eqnarray}
Here we assumed $L = 2 \times \text{odd}$ to be even.
$I_j$, $J_\alpha^n$, and ${J'}_\alpha^n$ are integer or half-odd
integer numbers, $N$ is the total number of electrons, $M' =
\sum_{n=1}^\infty n M_n'$, $\theta(x) = 2 \arctan(x)$, and
\begin{equation} \label{defthetas}
     \Theta_{nm} (x) = \left\{ \begin{array}{l}
	{\displaystyle
        \theta \left( \frac{x}{|n - m|} \right) +
        2 \theta \left( \frac{x}{|n - m| + 2} \right) + \cdots +
        2 \theta \left( \frac{x}{n + m - 2} \right) +
        \theta \left( \frac{x}{n + m} \right), \: \text{if} \quad
	n \ne m,} \\[3ex]
	{\displaystyle
        2 \theta \left( \frac{x}{2} \right) +
        2 \theta \left( \frac{x}{4} \right) + \cdots +
        2 \theta \left( \frac{x}{2n - 2} \right) +
        \theta \left( \frac{x}{2n} \right), \: \text{if} \quad n = m.}
	\end{array} \right.
\end{equation}
The branch of $\arcsin(x)$ in (\ref{t3}) is fixed as $- \pi/2 \le
\text{Re} (\arcsin (x)) \le \pi/2$. $M_n$ and $M_n'$ are the numbers of
$\Lambda$ strings of length $n$, and $\Lambda'$ strings of length $n$
in a specific solution of the system (\ref{t1})-(\ref{t3}). The
integer (half-odd integer) numbers in (\ref{t1})-(\ref{t3}) have
ranges
\begin{eqnarray} \label{r1}
     && - \frac{L - 1}{2} \le I_j \le \frac{L - 1}{2}, \\ \label{r2}
     && |J_\alpha^n| \le \frac{1}{2}
        \left(N - 2M' - \sum_{m=1}^\infty t_{nm} M_m - 1 \right), \\
	\label{r3}
     && |{J'}_\alpha^n| \le \frac{1}{2}
        \left(L - N + 2M' - \sum_{m=1}^\infty t_{nm} M_m' - 1 \right),
\end{eqnarray}
where $t_{mn} = 2 \min (m,n) - \delta_{mn}$. Each set of numbers
$\{I_j\}, \{J_\alpha^n\}, \{{J'}_\alpha^n\}$ is in one-to-one
correspondence with a set of rapidities $\{k_j\}, \{\Lambda_\alpha^n\},
\{{\Lambda'}_\alpha^n\}$, which in turn unambiguously specifies
one Bethe eigenstate of the Hubbard Hamiltonian. Thus the ground state
and all excited states can be constructed by specifying a set of
numbers $\{I_j\}, \{J_\alpha^n\}, \{{J'}_\alpha^n\}$ and then taking the
thermodynamic limit.

It is our aim to prove eq.\ (\ref{centre}) for the charge singlet
excitation over the half-filled ground state. The ground state is
characterized  by $N = L$, $M_1 = L/2$. In this case the inequalities
(\ref{r1}) and (\ref{r2}) lead to unique distributions of the quantum
numbers $I_j$ and $J_\alpha^1$,
\begin{equation} \label{ijg}
     I_j = - (L + 1)/2 + j \quad, \quad j = 1, \dots, L \quad, \quad
     J_\alpha^1 = - (L + 2)/4 + \alpha \quad, \quad
                \alpha = 1, \dots, L/2.
\end{equation}
Eqs.\ (\ref{t1}) and (\ref{t2}) reduce to
\begin{eqnarray} \label{g1}
     && L k_j = 2\pi I_j - \sum_{\alpha = 1}^{L/2}
                \theta \left( \frac{\sin k_j - \Lambda_\alpha}{U}
		       \right) \quad, \quad j = 1, \dots, L,\\
		       \label{g2}
     && \sum_{j=1}^L \theta \left( \frac{\Lambda_\alpha - \sin k_j}{U}
		       \right) = 2\pi J_\alpha^1 +
        \sum_{\beta = 1}^{L/2} \theta \left(
	   \frac{\Lambda_\alpha - \Lambda_\beta}{2U} \right)
	   \quad, \quad \alpha = 1, \dots, L/2.
\end{eqnarray}
In the thermodynamic limit eqs.\ (\ref{g1}) and (\ref{g2}) turn
into the well known integral equations \cite{LiWu68}
\begin{eqnarray} \label{gi1}
     \rho(k) & = & \frac{1}{2\pi} + \frac{1}{\pi} \cos k
                   \int_{- \infty}^\infty d \Lambda \;
		   \frac{U}{U^2 + (\sin k - \Lambda)^2}
		   \sigma(\Lambda), \\ \label{gi2}
     \sigma(\Lambda) & = & \frac{1}{2\pi^2} \int_{-\pi}^\pi dk \;
		   \frac{U}{U^2 + (\sin k - \Lambda)^2}
		   - \frac{1}{\pi} \int_{-\infty}^\infty d \Lambda' \;
		   \frac{2U}{4U^2 + (\Lambda - \Lambda')^2}
		   \sigma(\Lambda')
\end{eqnarray}
for the densities $\rho(k_j) = 1/(L(k_{j+1} - k_j))$ and
$\sigma(\Lambda_\alpha) = 1/(L(\Lambda_{\alpha + 1} - \Lambda_\alpha))$.

The charge singlet is  characterized  by $N = L$, $M_1 = L/2 - 1$ and
$M_1' = 1$ \cite{EsKo94a,EsKo94b}. Thus $M' = 1$, and the number of
unbound electrons is $L - 2M' = L - 2$. We will denote quantities which
describe the charge singlet by a tilde. Eqs.\ (\ref{r2}) and (\ref{r3})
uniquely determine the set $\{\tilde J_\alpha^1\}$ and the number
$\tilde {J'}^1$,
\begin{equation}
     \tilde J_\alpha^1 = - L/4 + \alpha \quad
                         \alpha = 1, \dots, L/2 - 1 \quad, \quad
     \tilde {J'}^1 = 0.
\end{equation}
The set $\{\tilde I_j\}$, however, is not uniquely determined by
the inequality (\ref{r3}). There are $L \choose 2$ inequivalent such
sets, which are parameterized  by two vacancies $I_1^h$ and $I_2^h$ in the
distribution (\ref{ijg}) of the numbers $I_j$, which characterize the
ground state. These two vacancies determine two charge rapidities
$k_1^h$ and $k_2^h$ via eqs.\ (\ref{g1}) and (\ref{g2}). In the
thermodynamic limit the charge rapidities densely fill the interval
$[-\pi,\pi]$, and $k_1^h$ and $k_2^h$ become the two free parameters of
the charge singlet excitation. We see already at this stage, that
there can not be a third free parameter, since $\tilde{J'}^1 = 0$ is
fixed. For the charge singlet excitation eqs.\ (\ref{t1})-(\ref{t3})
reduce to
\begin{eqnarray} \label{e1}
     && L \tilde k_j = 2\pi \tilde I_j - \sum_{\alpha = 1}^{L/2 - 1}
                \theta \left(
		       \frac{\sin \tilde k_j - \tilde \Lambda_\alpha}{U}
		       \right)
                - \theta \left( \frac{\sin \tilde k_j - \Lambda'}{U}
		       \right)
		       \quad, \quad j = 1, \dots, L - 2,\\
		       \label{e2}
     && \sum_{j=1}^{L -2} \theta \left(
                       \frac{\tilde \Lambda_\alpha - \sin \tilde k_j}{U}
		       \right) = 2\pi \tilde J_\alpha^1 +
        \sum_{\beta = 1}^{L/2 - 1} \theta \left(
	   \frac{\tilde \Lambda_\alpha -
	                             \tilde \Lambda_\beta}{2U} \right)
	   \quad, \quad \alpha = 1, \dots, L/2 - 1, \\ \label{e3}
     && L(\arcsin(\Lambda' + \text{i} U) +
                               \arcsin(\Lambda' - \text{i} U)) =
        \sum_{j=1}^{L-2} \theta \left(
	                \frac{\Lambda' - \sin \tilde k_j}{U} \right).
\end{eqnarray}
Let us subtract (\ref{g1}) from (\ref{e1}) for $I_l = \tilde I_j$,
$j = 1, \dots, L-2$, and (\ref{g2}) from (\ref{e2}) for $\alpha = 1,
\dots, L/2 - 1$. Then, taking the thermodynamic limit, we obtain
a pair of integral equations for the shift functions
\begin{equation}
     F^c (k_j) = \frac{\tilde k_j - k_j}{k_{j+1} - k_j} \quad, \quad
     F^s (\Lambda_\alpha) = \frac{\tilde \Lambda_\alpha -
        \Lambda_\alpha}{\Lambda_{\alpha + 1} - \Lambda_\alpha}.
\end{equation}
These integral equations are
\begin{eqnarray} \label{f1}
     F_{CS}^c (k) & = & - \frac{1}{2} - \frac{1}{2\pi}
                        \theta \left( \frac{\sin k - \Lambda'}{U}
			\right)
			+ \frac{1}{\pi} \int_{-\infty}^\infty
			d \Lambda \;
			\frac{U}{U^2 + (\sin k - \Lambda)^2}
			F_{CS}^s (\Lambda), \\ \label{f2}
     F_{CS}^s (\Lambda) & = & 1 + \frac{1}{2\pi} \sum_{l = 1}^2
                              \theta \left(
			      \frac{\Lambda - \sin k_l^h}{U} \right) -
			      \frac{1}{\pi} \int_{-\infty}^\infty
			      d \Lambda' \;
			      \frac{2U}{4U^2 + (\Lambda - \Lambda')^2}
			      F_{CS}^s (\Lambda').
\end{eqnarray}
Here we have supplied an index ``CS'' to  emphasize  that we are dealing
with the charge singlet excitation. The solution $F_{CS}^c (k)$ of
(\ref{f1}) can be found on page 526 of \cite{EsKo94b}. It is a single
valued function of $\sin k$. In the derivation of (\ref{gi2}) and
(\ref{f2}) we made use of the following elementary lemma,
\begin{equation} \label{lemma}
     \int_{-\pi}^\pi dk \; f(\sin k) \cos k = 0,
\end{equation}
which holds for arbitrary single valued functions $f(x)$. This lemma
follows from the identities $\sin (\pi - k) = \sin k$ and
$\cos (\pi - k) = - \cos k$.

Note that there are still three free parameters, $k_1^h$, $k_2^h$
and $\Lambda'$, in (\ref{f1}) and (\ref{f2}). $\Lambda'$ becomes
fixed as a function of $k_1^h$ and $k_2^h$ by considering eq.\
(\ref{e3}), which in the thermodynamic limit turns into
\begin{eqnarray} \nonumber
     \lefteqn{
     L(\arcsin(\Lambda' + \text{i}U) + \arcsin(\Lambda' - \text{i}U))}
     \\ & = &
     L \int_{-\pi}^\pi dk \; \theta \left( \frac{\Lambda' - \sin k}{U}
        \right) \rho (k) - 2 \int_{-\pi}^\pi dk \;
	\frac{U}{U^2 + (\Lambda' - \sin k)^2} F_{CS}^c (k) \cos k
	- \sum_{l=1}^2 \theta \left(\frac{\Lambda' - \sin k_l^h}{U}
	\right) + {\cal O} \left( \frac{1}{L} \right) \nonumber
     \\ & = & \label{rhs}
     \frac{L}{2\pi} \int_{-\pi}^\pi dk \;
        \theta \left( \frac{\Lambda' - \sin k}{U} \right)
	- \sum_{l=1}^2 \theta \left(\frac{\Lambda' - \sin k_l^h}{U}
	\right) + {\cal O} \left( \frac{1}{L} \right).
\end{eqnarray}
In oder to obtain the second of the above equalities we have used
(\ref{gi1}), (\ref{f1}) and the lemma (\ref{lemma}). Let us calculate
the integral
\begin{equation}
     I (\Lambda') = \frac{1}{2\pi} \int_{-\pi}^\pi dk \;
                          \theta \left( \frac{\Lambda' - \sin k}{U}
			  \right) .
\end{equation}
on the right hand side of (\ref{rhs}). First note that $I(0) = 0$.
The derivative of $I (\Lambda')$ can be represented as
\begin{equation} \label{integral}
     I'(\Lambda') = \frac{1}{\pi} \int_{-\pi}^\pi dk \;
                              \frac{U}{U^2 + (\Lambda' - \sin k)^2}
                        = \text{Re} \left\{ \frac{1}{2\pi \text{i}}
			      \oint dz \frac{4}{z^2 + 2z(U - \text{i}
			      \Lambda') - 1} \right\} ,
\end{equation}
where the contour of integration is the unit circle. Let
\begin{equation}
     p(z) = z^2 + 2z(U - \text{i} \Lambda') - 1 = (z - z_1)(z - z_2).
\end{equation}
We see that the poles of the integrand in (\ref{integral}) are related
as
\begin{equation} \label{poles}
     z_1 = - 1/z_2.
\end{equation}
Thus, only one of these poles, say $z_1$, lies inside the unit circle.
Using (\ref{poles}) we obtain $I'(\Lambda')$ as a function of
$z_1$,
\begin{equation} \label{phiprime}
     I'(\Lambda') = 2 \text{Re} \left\{
                          \frac{2}{z_1 + z_1^{-1}} \right\} .
\end{equation}
Let us parameterize  the pole at $z_1$ as $z_1 = e^{\text{i} \alpha}$.
Since $z_1$ is located inside the unit circle, $\text{Im} (\alpha) > 0$.
Using $p(z_1) = 0$ we find that
\begin{equation}
     \sin \alpha = \Lambda' + \text{i} U \quad,
                   \quad \text{Im} (\alpha) > 0.
\end{equation}
This equation fixes $\alpha$ modulo $2\pi$. Now $U > 0$ and $\text{Im}
(\alpha) > 0$, and therefore
\begin{equation}
     \cos \text{Re} (\alpha) = \frac{U}{\sinh \text{Im} (\alpha)} > 0.
\end{equation}
We conclude that $- \pi/2 < \text{Re} (\alpha) < \pi/2$. Thus (see the
definition below (\ref{defthetas})) $\alpha = \arcsin (\Lambda' +
\text{i} U)$.  Integrating (\ref{phiprime}) with respect to $\Lambda'$
and using $I (0) = 0$ to fix the integration constant we arrive at
\begin{equation}
     I(\Lambda') = 2 \text{Re} (\alpha)
                       = 2 \text{Re} (\arcsin(\Lambda' + \text{i}U))
		       = \arcsin(\Lambda' + \text{i} U) +
		         \arcsin(\Lambda' - \text{i} U).
\end{equation}
We may now insert this result into eq.\ (\ref{rhs}). This yields
\begin{equation}
     \theta \left(\frac{\Lambda' - \sin k_1^h}{U} \right) =
     - \theta \left(\frac{\Lambda' - \sin k_2^h}{U} \right).
\end{equation}
Dividing by 2 and taking $\tan$ gives
\begin{equation}
     2 \Lambda' = \sin k_1^h + \sin k_2^h.
\end{equation}
So we have accomplished our task to show that (2) follows from eqs.\
(20)-(22) of \cite{BrAn98}.
\section{}
Let us show that for one $k$-$\Lambda$ string (and several real $k_j$
and real $\lambda_\gamma$) eq.\ (7) of \cite{BrAn98} follows from
eqs.\ (5), (6), (8) of \cite{BrAn98}. In case of a single
$k$-$\Lambda$-two string the numbers $N^u$ and $M^u$ of \cite{BrAn98}
are $N^u = N - 2$, $M^u = M - 1$, where $N$ is the total number of
electrons and $M$ is the total number of down spins. There are $N - 2$
real charge rapidities $k_1, \dots, k_{N-2}$, which correspond to
unbound particles. The string is  characterized  by two charge rapidities
$k^+$ and $k^-$ and by one spin rapidity $\Lambda$. $k^\pm$ and
$\Lambda$ satisfy
\begin{eqnarray}
     \sin k^+ = \Lambda - \text{i} u/4 + \varepsilon^+ \, \\
     \sin k^- = \Lambda + \text{i} u/4 - \varepsilon^- \,
\end{eqnarray}
where $\varepsilon^+$ and $\varepsilon^-$ become small as the length
$L$ of the lattice becomes large. Eqs.\ (5)-(8) of \cite{BrAn98} are
\begin{eqnarray} \label{5}
     e^{\text{i} k_j L} & = & \frac{\Lambda - \sin k_j - \text{i} u/4}
                                   {\Lambda - \sin k_j + \text{i} u/4}
                     \prod_{\delta = 1}^{M-1}
                     \frac{\lambda_\delta - \sin k_j - \text{i} u/4}
                          {\lambda_\delta - \sin k_j + \text{i} u/4},
                     \quad j = 1, \dots, N - 2 , \\ \label{6}
     \prod_{\delta = 1 \atop \delta \ne \gamma}^{M-1}
        \frac{\lambda_\gamma - \lambda_\delta - \text{i} u/2}
             {\lambda_\gamma - \lambda_\delta + \text{i} u/2}
     & = & \prod_{j=1}^{N-2}
           \frac{\lambda_\gamma - \sin k_j - \text{i} u/4}
                {\lambda_\gamma - \sin k_j + \text{i} u/4},
                     \quad \gamma = 1, \dots, M - 1 , \\ \label{7}
     \prod_{\delta = 1}^{M-1}
        \frac{\Lambda - \lambda_\delta - \text{i} u/2}
             {\Lambda - \lambda_\delta + \text{i} u/2}
     & = & \frac{\varepsilon^+}{\varepsilon^-} \prod_{j=1}^{N-2}
           \frac{\Lambda - \sin k_j - \text{i} u/4}
                {\Lambda - \sin k_j + \text{i} u/4}, \\ \label{8}
     e^{\text{i}(k^+ + k^-)L}
        & = & \frac{\varepsilon^+}{\varepsilon^-}
	      \prod_{\delta = 1}^{M-1}
              \frac{\lambda_\delta - \Lambda - \text{i} u/2}
                   {\lambda_\delta - \Lambda + \text{i} u/2}.
\end{eqnarray}
The ratio
\begin{equation}
     \frac{\varepsilon^+}{\varepsilon^-} = - \;
        \frac{\Lambda - \sin k^+ - \text{i} u/4}
             {\Lambda - \sin k^- + \text{i} u/4}
\end{equation}
is denoted by $e^{\text{i} \varphi}$ in \cite{BrAn98}. We want to show
that (\ref{7}) algebraically follows from (\ref{5}), (\ref{6}) and
(\ref{8}). First note that momentum conservation implies that
\begin{equation} \label{momentum}
     \exp \left({\textstyle \text{i} L \left(
        \sum_{j=1}^{N-2} k_j + k^+ + k^-
        \right)} \right) = 1.
\end{equation}
Multiplying all eqs.\ (\ref{6}) we get
\begin{equation} \label{prodisone}
     \prod_{j=1}^{N-2} \prod_{\delta = 1}^{M-1}
        \frac{\lambda_\delta - \sin k_j - \text{i} u/4}
             {\lambda_\delta - \sin k_j + \text{i} u/4} =
     \prod_{\delta = 1}^{M-1}
        \prod_{\gamma = 1 \atop \gamma \ne \delta}^{M-1}
	\frac{\lambda_\delta - \lambda_\gamma - \text{i} u/2}
	     {\lambda_\delta - \lambda_\gamma + \text{i} u/2} = 1.
\end{equation}
Now let us multiply all eqs.\ (\ref{5}). Taking into account
(\ref{prodisone}) we get
\begin{equation}
     e^{\text{i}L \sum_{j=1}^{N-2} k_j} =
        \prod_{j=1}^{N-2} \frac{\Lambda - \sin k_j - \text{i} u/4}
                               {\Lambda - \sin k_j + \text{i} u/4}.
\end{equation}
Let us multiply this by (\ref{8}) and use (\ref{momentum}). Then
\begin{equation}
     1 = \exp \left({\textstyle \text{i} L \left(
         \sum_{j=1}^{N-2} k_j + k^+ + k^-
         \right)} \right) =
         \frac{\varepsilon^+}{\varepsilon^-} \prod_{\delta = 1}^{M-1}
         \frac{\lambda_\delta - \Lambda - \text{i} u/2}
              {\lambda_\delta - \Lambda + \text{i} u/2}
         \prod_{j=1}^{N-2}
	 \frac{\Lambda - \sin k_j - \text{i} u/4}
	      {\Lambda - \sin k_j + \text{i} u/4},
\end{equation}
which is equivalent to (\ref{7}) (or (7) in \cite{BrAn98}).


\end{document}